\documentclass[sigconf]{acmart}
\usepackage{framed}

\AtBeginDocument{%
  \providecommand\BibTeX{{%
    \normalfont B\kern-0.5em{\scshape i\kern-0.25em b}\kern-0.8em\TeX}}}

\copyrightyear{2024}
\acmYear{2024}
\setcopyright{rightsretained}
\acmConference[CHIIR '24]{Proceedings of the 2024 ACM SIGIR Conference on Human Information Interaction and Retrieval}{March 10--14, 2024}{Sheffield, United Kingdom}
\acmBooktitle{Proceedings of the 2024 ACM SIGIR Conference on Human Information Interaction and Retrieval (CHIIR '24), March 10--14, 2024, Sheffield, United Kingdom}\acmDOI{10.1145/3627508.3638298} \acmISBN{979-8-4007-0434-5/24/03}

\usepackage[inline]{enumitem}
\usepackage{cleveref}
\usepackage{color}
\usepackage{geometry}
\definecolor{lightgray}{gray}{0.9}

\newsavebox{\promptbox}

\usepackage{booktabs}
\usepackage{siunitx}
\usepackage{float}

\begin{document}

\title[BibSonomy Meets ChatLLMs for Publication Management]{From Chat to Publication Management: Organizing your related work using BibSonomy \& LLMs}

\author{Tom Völker}
\email{tom.voelker@informatik.uni-wuerzburg.de}

% TODO: More Info??
\affiliation{%
  \institution{Center for Artificial Intelligence and Data Science (CAIDAS), Institute for Computer Science, University of Würzburg}
  \city{Würzburg}
  \country{Germany}
  \postcode{97074}
}

\author{Jan Pfister}
\email{pfister@informatik.uni-wuerzburg.de}

% TODO: More Info??
\affiliation{%
  \institution{Center for Artificial Intelligence and Data Science (CAIDAS), Institute for Computer Science, University of Würzburg}
  \city{Würzburg}
  \country{Germany}
  \postcode{97074}
}

\author{Tobias Koopmann}
\email{koopmann@informatik.uni-wuerzburg.de}

\affiliation{%
  \institution{Center for Artificial Intelligence and Data Science (CAIDAS), Institute for Computer Science, University of Würzburg}
  \city{Würzburg}
  \country{Germany}
  \postcode{97074}
}

\author{Andreas Hotho}
\email{hotho@informatik.uni-wuerzburg.de}
\affiliation{%
  \institution{Center for Artificial Intelligence and Data Science (CAIDAS), Institute for Computer Science, University of Würzburg}
  \city{Würzburg}
  \country{Germany}
  \postcode{97074}
}

\ccsdesc[500]{Human-centered computing~Natural language interfaces}
\ccsdesc[500]{Human-centered computing~Web-based interaction}
\ccsdesc[500]{Applied computing~Document management}
\ccsdesc[200]{Applied computing~Digital libraries and archives}
\ccsdesc[100]{Applied computing~Document metadata}
\ccsdesc[500]{Information systems~Search interfaces}
\ccsdesc[400]{Information systems~Application servers}

\begin{abstract}

The ever-growing corpus of scientific literature presents significant challenges for researchers with respect to discovery, management, and annotation of relevant publications. 
Traditional platforms like Semantic Scholar, BibSonomy, and Zotero offer tools for literature management, but largely require manual laborious and error-prone input of tags and metadata. 
Here, we introduce a novel retrieval augmented generation system that leverages chat-based large language models (LLMs) to streamline and enhance the process of publication management.
It provides a unified chat-based interface, enabling intuitive interactions with various backends, including Semantic Scholar, BibSonomy, and the Zotero Webscraper. 
It supports two main use-cases: (1)  Explorative Search \& Retrieval - leveraging LLMs to search for and retrieve both specific and general scientific publications, while addressing the challenges of content hallucination and data obsolescence; and 
    (2)  Cataloguing \& Management - aiding in the organization of personal publication libraries, in this case BibSonomy, by automating the addition of metadata and tags, while facilitating manual edits and updates.
We compare our system to different LLM models in three different settings, including a user study, and we can show its advantages in different metrics.  
\end{abstract}

\keywords{
% BibSonomy,
ChatGPT,
Publication Management,
RAG,
% Conversational Interface,
Academic Search
}

% TODO
%\received{31 October 2023}
%\received[accepted]{12 December 2023}

\maketitle

\begin{figure*}
    \centering
    \includegraphics[width=0.8\textwidth]{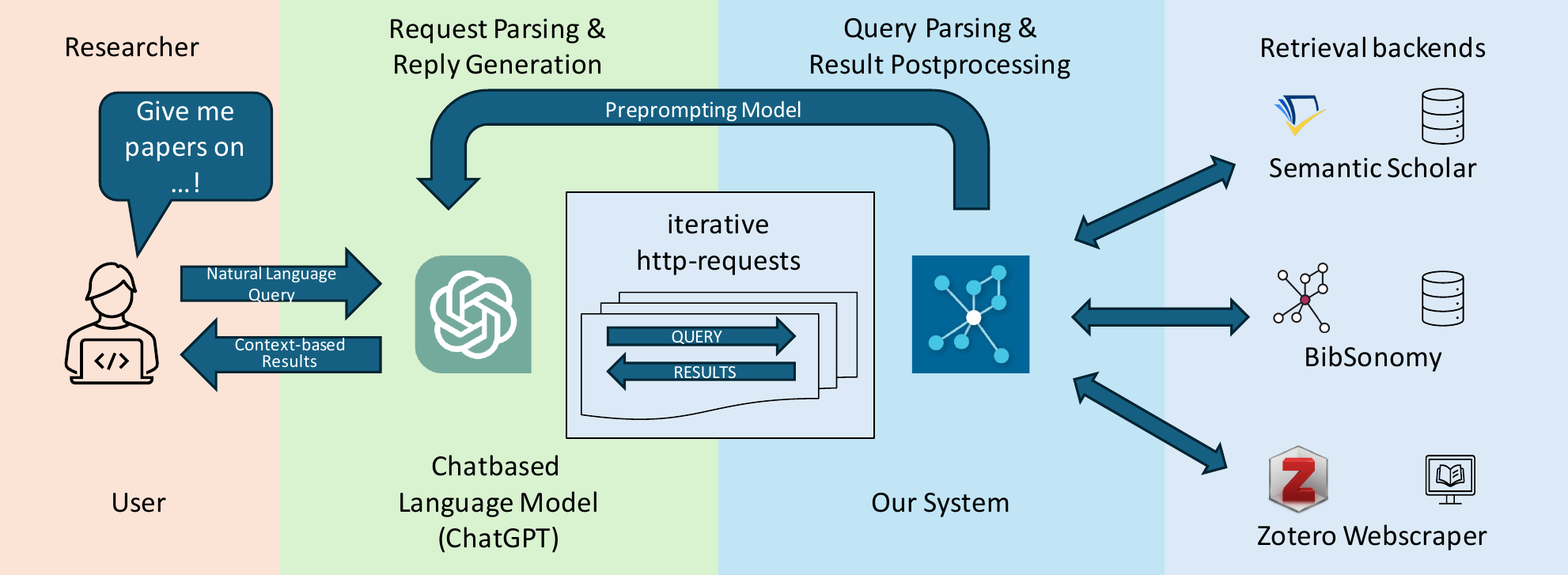}
    \caption{Conceptual overview of our plugin. 
    On the left-hand side, the user inserts queries into a chat-based language model. 
    The model then decides whether to use the plugin.
    If so, it queries the plugin management server.
    This server then communicates with different back-end APIs, post-processes their answers, and returns the collected results back to the chat-based language model.
    With this additional information, the model generates a reply for the user.
    }
    \Description{Figure 1 presents a conceptual overview of a plugin system integrated with a chat-based language model. On the left side, a researcher (user) inputs a natural language query such as ``Give me papers on ...!'' to the system. The user is represented as a stick figure next to a speech bubble with code brackets, indicating a query is being made. The query is then processed by the chat-based language model (ChatGPT), depicted as a brain-like icon. This model forwards the query to the plugin management server shown as a box labeled ``Our System,'' which performs iterative http-requests, symbolized by arrows looping between the ``QUERY'' and ``RESULTS'' inside the box. The system then communicates with various retrieval backends (Semantic Scholar, BibSonomy, Zotero Webscraper), each represented by its respective logo, to gather results. These results are post-processed and returned to the language model, which generates a reply for the user. The process flow is demonstrated through arrows indicating the direction of data transfer between components.}
    \label{fig:figure1}
\end{figure*}

\section{Introduction}
The constantly expanding scientific landscape makes it a significant challenge for researchers to efficiently find, manage and keep track of relevant literature.
To this end, various platforms, such as Semantic Scholar~\cite{SemanticScholar}, BibSonomy~\cite{bibsonomy}, or Zotero~\cite{zotero} offer a range of tools to search, categorize, and index these publications.
Unfortunately, manual and consistent addition of metadata and tags can be error-prone and time-consuming.

Fortunately, recent advances in chat-based large language models offer a promising opportunity to support and streamline this publication management process.
In this context, we introduce a retrieval augmented generation system tailored for scientific research and publication management.
This system is designed to enable the user to seamlessly interact with various backends, via an intuitive chat interface.
Thus, it provides a simple, cohesive, and intuitive interface, enabling the user to interact with different platforms using natural language. 
Currently, three backends are supported: SemanticScholar, BibSonomy, and the Zotero Webscraper.
Two separate but closely related tasks are accomplished by our system: seeking out new or known scientific publications, and managing a personal virtual library of self-posted literature.

{\textbf{(Explorative) Search \& Retrieval}}\,
This use case involves searching for specific academic literature that the user wants to retrieve or conducting a broader, more general search when the sought-after material is not yet known to the user.
To accomplish this, we utilize recent advances in chat-based large language models.
These pretrained models have been shown to possess vast amounts of general knowledge, enabling them to confidently describe complex concepts~\cite{DBLP:conf/nips/BrownMRSKDNSSAA20}, reason over difficult problems~\cite{DBLP:conf/ijcai/ClarkTR20}, and can even provide details on scientific publications included in their training corpus (see \ref{sec:eval}).
However, relying solely on these models for publication management can be problematic or flat out impossible due to numerous serious issues arising with that:
First, the tendency of the models to hallucinate wrong, but plausible sounding content~\cite{hallucinations,DBLP:conf/fat/BenderGMS21,DBLP:conf/acl/LinHE22,DBLP:journals/corr/abs-2107-03374} and second, the currency of these models, as they are only as up-to-date as the data on which they are trained.
Thus, these models cannot return the most recent publication and may even invent publications that do not exist.
In contrast, our system enables the model to query different knowledge bases to provide well-founded and verifiable answers.

{\textbf{Cataloguing \& Management}}\, The second use-case centers around efficiently managing a researcher's personal publication library.
This includes not only the organization and collection of publications but also the addition of correct metadata and, ideally, relevant and user-specific tags and descriptions.
Although metadata can be collected from large-scale academic sources such as Semantic Scholar, manually transferring or even updating existing entries with new information, such as PDFs or a newly relevant, user-specific tag is usually tedious work.
To mitigate this, our system allows for editing of posts, allowing for manual addition of metadata or automatic collection of data from other sources.

In this work, we introduce an open-source backend server\footnote{\url{https://bitbucket.org/bibsonomy/bibsonomy-llm-scientific-retrieval-plugin}} that enables large-language models (LLMs) to query different services.
This server can be used as a retrieval engine for LLMs like ChatGPT. 
It allows to exploratively search for related work leveraging the open-world knowledge in LLMs, and organize self-managed publications by automating the process of adding relevant tags and metadata to publications. 
Additionally, we evaluate our tool in three settings, including a user study compared to other tools in this domain.

\section{Related Work}

Large language models with transformer architectures~\cite{bert,xlnet} represent a major advance in natural language processing, offering a multitude of potential applications~\cite{biobert,DBLP:conf/nips/BrownMRSKDNSSAA20,DBLP:conf/ijcai/ClarkTR20}. 
However, a significant challenge associated with these models is their tendency to generate inaccurate or fabricated information~\cite{DBLP:conf/fat/BenderGMS21,DBLP:conf/acl/LinHE22,DBLP:journals/corr/abs-2107-03374}. 
To mitigate this problem and improve the reliability of these models, recent research efforts are focused on the integration of structured data.~\cite{toolformer,RAG} 
This aims to provide these models with structured information by providing viable facts and help the models generate accurate answers.

\section{BibSonomy Chat-LLM Service}
Our open-source backend service now builds on this concept and addresses the settings of \textit{Search \& Retrieval} as well as \textit{Cataloguing \& Management}.
It enables chat-based large language models (LLMs) to communicate with different bibliography services, as depicted in \Cref{fig:figure1}.
This approach follows the ``toolformer''-paradigm~\cite{toolformer}, which allows large language models to proficiently control and interact with different services.
In this setting, the LLM effectively functions as a mediator between unstructured, free-form user input and structured queries to a backend system.
User queries can range from simple tasks such as converting an APA-formatted citation into bibtex (Appendix~\ref{app:parse2bib}) to more intricate, multi-step requests like locating the Llama 2 Foundation Paper and adding it to the user's BibSonomy Account (Appendix~\ref{app:add2BibSonomy}). 
% Here, user queries can consist of simpler tasks such as ``parse this APA-formatted citation into bibtex'' (\Cref{app:parse2bib}), or more complicated multi-step requests such as ``Can you please find the Llama 2 Foundation Paper for me and add it to my BibSonomy Account? Please add a suitable description and tags including the tag ``paper\_demo''.~(\Cref{app:add2BibSonomy}).
For this demo, we deploy our service within the OpenAI-LLM-plugin ecosystem~\cite{OpenAIPlugins}, but want to emphasize that any other pretrained and promptable (e.g. self-hosted) LLM can be used with our system~\cite{OpenAgents}.

\subsection{Prompting the Model}
We first describe how the model is prompted and which information is provided to the model.
Technically, the entire interface is defined within a single \texttt{yaml}-file given to the model before the user interacts with it.
Here, all available http-endpoints, including their respective parameters, are defined and explained to the model using natural language.
The endpoints we offer are implemented in an intermediate Python server, which serves as a middle layer between the different backends and the chat model (\Cref{fig:figure1}).
After receiving the response from our service, the LLM will be prompted with this information to generate a more concise response for the user. 

Furthermore, an initial system prompt is given to make the model aware of its environment (Appendix~\ref{app:model_desc}).
In addition to providing a brief overview of the system's functionality and target use cases, this prompt primarily serves to introduce the model to the desired mode of operation.
On the one hand, the prompt explicitly explains how good search queries should be formulated, but it also points out the relevance of data integrity in the context of scientific work (to minimize hallucinations).
Furthermore, it explains the desired procedure when posting content to BibSonomy (e.g.\ how tags and descriptions should be used, etc.).
Finally, we have had great success including additional system prompts within the API responses.
%\footnote{also used by the Wikipedia plugin \url{https://gitlab.wikimedia.org/repos/machine-learning/chatgpt-plugin}}.
This approach adds appropriate suggested actions or hints to the model based on the API endpoint used (and other parameters) to make the model behave as desired.
%Using this method, the factuality of the model can be improved, method calls can be optimized, and saving publications to BibSonomy can be facilitated.

\subsection{Endpoints Provided by the Backend}
{\textbf{(Explorative) Search \& Retrieval}}\,
The first use-case revolves around exploring established or undiscovered scientific work, for which we offer two simple endpoints to the model.
They allow querying different backends and using the results to generate an answer, inspired by the ``Retrieval Augmented Generation''-paradigm~\cite{RAG}.
The two provided http-endpoints are called \texttt{/search} and \texttt{/details}, which can be used with a multitude of optional query parameters.
They provide a unified and simple access to an easily extendable list of backends (currently: BibSonomy, SemanticScholar, and Zotero scrapers).
Support can be easily extended to other back-ends, given they provide API endpoints conceptually matching either of these endpoints.
\textit{The ``/search''-endpoint} is designed to handle a variety of query inputs, accepting free-form text queries or parameters such as title, authors, or keywords. 
Upon receiving a request, the service transforms it internally into native query formats compatible with various supported backends, specifically for this endpoint, BibSonomy and SemanticScholar.
To mimic a researcher's approach, the model can also enrich the search by including supplementary search terms closely related to the main query, allowing for a more holistic understanding of the topic.
By default, both backends are queried for the top results for each query, with the results then getting merged and re-ranked within our backend, before they are returned to the chat model.
We concatenate the results from both backends and reorder them according to the results BM25 scores~\cite{BM25}. 
Here, the system considers the query type and the number of platforms on which a result appears. 
In particular, results sourced from the primary query are given substantially more weight than those from supplementary queries, ensuring the relevance of returned results.
The model can also choose to explicitly query a single backend or search for a different number of results by specifying the query parameters. 
Given the limited context length of LLMs~\cite{xlnet}, they cannot handle an extremely long input. 
Extensive data, such as numerous relevant publications, returned by the backend may lead to struggles with the response length.
We thus provide two parameters, which are by default set to be rather strict and have to be explicitly raised by the model to prevent it from being quickly overwhelmed by long results: (1)  the number of results returned, and
    (2) the granularity of the returned information, which is ``basic'' by default (title, author, year, number of citations) but can be specified to be more ``verbose'', containing more metadata information provided by the respective platform.
Furthermore, a unique platform-specific identifier is returned for each publication, providing a short handle to the model to further reference the specific publication when communicating with our backend.
This has two benefits:
    (1)  makes it easier for the model to clearly reference a result with only comparatively few tokens added to the context window,    (2)  while also speeding up the interaction as less tokens have to be generated when formulating the requests that are sent to either endpoint.
For the \textit{``/details''-endpoint} this unique identifier is enough to retrieve all metadata information provided by the platform.
Furthermore, it can also resolve various other commonly used identifiers such as DOIs, arXiv-IDs, ACL-IDs, etc.\ , or even arbitrary URLs, which are then resolved using the well-maintained Zotero WebScraping server~\cite{ZoteroTranslationServer}.
This enables unique workflows, like providing a URL to a publication, for which the model then fetches basic metadata using Zotero, looks up the correct bibtex using Semantic Scholar, and posts it with all gathered metadata to the users' personal BibSonomy library (appendix~\ref{app:add2BibSonomyZotero}).

{\textbf{Publication Management}}\,
Once a publication is in the user's personal BibSonomy library, it can be further managed either ``traditionally'' via the BibSonomy interface, or again via this system using the chat interface.
This expands the model's capabilities to go beyond a simple ``read-only'' mode:
Here \textit{the ``/search''-endpoint} provides the ability to search within the users' private library, either filtered by keywords, or manually assigned keyword tags.
Furthermore, it is possible to edit posts, for example, to add or update bibliography metadata, change user-given tags, or even upload an associated PDF file.
Since our backend offers access to the user's previously assigned tags, the LLM is capable of comprehending the user-specific tagging system that has already been employed. 
Consequently, it can accurately apply tags to existing or newly added literature in a manner consistent with the user's approach, utilizing pre-existing tags or introducing new ones as needed and appropriate.
Furthermore, when adding a new publication, the model can add a short description to the publication, briefly describing the content or relevant context of the publication.
Whenever a publication is added or modified using a language model, it is automatically assigned the tags \texttt{posted\_with\_chatgpt} or \texttt{edited\_with\_chatgpt}, thus the users are aware that LLMs have been used in the processing of this entry.

\section{Evaluation}
\label{sec:eval}
Our system has been successfully installed and used by more than 350 unique users. 
Now, we will evaluate our chat integration by comparing it qualitatively and quantitatively with similar tools, namely
   (1)  Vanilla ChatGPT-4   (2) You.com~\cite{you}    (3)  ChatGPT-4 with Bing   (4) ChatGPT-4 with ScholarAI~\cite{ScholarAI}
    (5)  and finally ChatGPT-4 with BibSonomy (ours).
While some systems, like ScholarAI and ours, cater to academic research, others, such as Bing and You.com, rely on expensive internet browsing. 
Vanilla ChatGPT-4 relies solely on its training data and is used as a baseline. 
ScholarAI, sharing a similar concept with our plugin, is our main point of comparison. 
However, it focuses more on detailed data for specific papers than on an overall topic perspective. 
Moreover, their publication management options only extend to a rudimentary Zotero integration, requiring manual API key inputs. 
Additionally, ScholarAI is not open source, imposing paywalls and limits on searches, and lacks transparency on its processes and data.
We evaluate our backend's search and retrieval capabilities under three different aspects: user opinion, determinism of query results, and inference time.

{\textbf{User Studies:}}\,
We prompt each of the five systems introduced above with the same seven queries (appendix~\ref{app:evalprompts}) and ask 17 machine learning PhD candidates to rank the results against each other.
The prompts were carefully generated and selected by us beforehand to cover different levels of complexity and research approaches and focus on retrieving specific articles based on topics and authors (Appendix~\ref{app:evalprompts}).
Overall 219 votes were cast between the replies of two random models to the identical prompt\footnote{The website used, as well as the code for the evaluation, as well as all model prompts and replies can be found in the repository.}.
The decision is a blind (i.e.\ the experts do not know which LLM created which answer), pairwise comparison polling three different aspects:
Which response fits \textbf{intuitively better} for the query, is more \textbf{scientifically valid} (i.e.\ no hallucinations) and which response is more \textbf{up to date}.
From \Cref{tab:user_eval} we find that the answers provided by our system are clearly favored over all other responses (favored in about 90\% of the pairwise comparisons) in the three aspects evaluated.
ChatGPT4 with ScholarAI, as well as Bing, is closest to our system, as it is preferred in 53\% to 59\% of comparisons regarding how ``intuitive'' the result matches the user prompt and the ``validity'' of the reply.
Furthermore, the former is preferred in 73\% of the cases for ``recency'', while the latter is preferred only in 55\% of the comparisons here.
The other systems are generally preferred in less than 40\% of pairwise comparisons. 
Note that the metrics exceed 100\% due to the cumulative effect of pairwise comparisons across the five models, resulting in an aggregated total that surpasses the conventional percentage threshold.

%Vanilla GPT-4 exhibits strength when addressing topics within its training data, as demonstrated by its on-par results when asked about the XL-Net publication from year 2019 (\Cref{app:xlnet}).
%However, its aptitude diminishes for data from the current or previous year, underscored by its low ``recency'' score of 10\%, emphasizing the need for systems with retrieval backends.
The queries used can be divided into two categories: topic-related prompts and author-related prompts. 
For topic-related prompts, models with unlimited internet access, such as the Bing model, demonstrate a strong performance. 
Their strength lies in the ability to search the Internet, accessing diverse sources such as blogs and other non-traditional academic platforms. 
Yet, in direct comparisons, our system still emerges as the preferred choice, highlighting its proficiency in focused topic searches.

When considering author-specific prompts, the landscape changes.
Here, the breadth of information accessed by these internet-dependent models can sometimes work against them, leading to less coherent outputs when handling expansive data requests. 
While Scholar AI, as another database-reliant system, presents commendable results, our plugin consistently emerges as the top choice.
This highlights the benefit of using a backend with structural data access, which emphasizes the benefit of our system approach.

% When comparing our system to models with unlimited internet access, our system consistently stands out as the preferred choice.
% It is the dominant system in most comparisons, especially in searches that are very focused on a specific topic, with the Bing model trailing closely. 
% Its ability to scrape the Internet provides it with unique background details that are not usually accessible through standard academic data sources. 
% However, this approach leads to it regularly faltering in delivering a coherent understanding, especially with expansive data requests like our author-based searches.
% In these author-specific queries, while Scholar AI presented strong results, our system still kept the advantage.
% This highlights the benefit of utilizing a backend with structural data access, which emphasizes the benefit of our system approach.

{\textbf{Determinism of Query Results}}
Secondly, we evaluated reply determinism of different systems across several executions with an identical prompt. 
We argue that this is a core requirement of a reliable retrieval system, as it should not return a different random subset of relevant publications every time the user queries.
%Vanilla GPT-4 was left out here since all tested queries demanded information not contained in its training corpus.
For this assessment, we compare the four systems with access to external resources in two settings.
We prompt the models to retrieve three recent and three highly influential publications of a specific author (Appendix~\ref{app:hotho}), as well as identifying the Palm2~\cite{anil2023palm} publication, as ``a new model by one of the original BERT authors'' (Appendix~\ref{app:palm}).
We evaluated the number of identical linked publications for the same query in five iterations .
%The \texttt{BibSonomy} model consistently produced the exact same results in every run, distinguishing itself from other models.
Here, the BibSonomy plugin again consistently returns the correct publication in every run, showcasing the ability to access structured data and a coordinated search strategy.
Although ScholarAI's model is capable of consistently citing the most influential items, it only once provided the desired combination of both recent and influential publications and failed to recognize the Palm2 publication once.
On the contrary, GPT-4 with Bing and You.com showed general inconsistencies in their responses.
Even when browsing similar sources during its search, the final responses vary widely, such that instead of returning six publications as requested, the model frequently returns CVs and other unrelated information, often omitting the desired publications.
Both never correctly identified the Palm2 publication and often erroneously returned nonrelated work.
You.com was unable to provide consistent responses and additionally regularly returns non-existent publications and other hallucinations. 

{\textbf{Inference Time}}\,
Finally, we measure the average time-to-response of the different engines for the two prompts mentioned above.
Here, the time from the ``send'' action was measured until the model started to respond. %, to avoid penalizing longer and more detailed responses.
Vanilla GPT-4 and You.com were excluded based on their subpar response quality, as they were preferred in less than 50\% of the user evaluation instances.
While the BibSonomy and ScholarAI systems yield results within a similar time frame, GPT-4 with Bing requires on average more than three times longer to retrieve its data, occasionally exceeding two minutes before answering.
We argue that this significantly affects the usability of the model and shows the need for a use-case-tailored search backend.

% \textit{Thoughts:} \\
% - Grundsätzliches Problem: Aktuell sehr krass in der entwicklung\\
% - Gibt noch andere ChatGPT Plugins die während des Entwicklungsprozesses erschienen sind (hier muss man dann echt mal gucken wie man das am sinnvollsten macht. Ist halt echt einiges, aber auch echt einiges an crap). \\
% - Allgemein kann man sagen: Definitiv ist die Kombi aus Reserach und kostenlosem/open source Publikationsmanagement aktuell einzigartig \\
% Was man mal noch gucken sollte: \\
% - Kann man die Plugins irgendwie gegeneinander "antreten" lassen und wenn ja, kommen wir da gut weg? \\
% - Welche Sachen gibts außerhalb von ChatGPT noch? -> In jedem Fall ist das dann aber schon nicht mehr vergleichbar, weils ja gerade auch irgendwie um den "lightweight" approach geht

\section{Conclusion}

We present the BibSonomy Chat-LLM Service, which streamlines and simplifies publication management for researchers. 
The plugin accesses different back-ends to conduct smart searches for new, related work from large data sources like Semantic Scholar.
Furthermore, it seamlessly integrates with the BibSonomy publication management system, enabling it to add, edit, and fetch data from each user's personal collection. 
By using BibSonomy's tagging system, the plugin can intelligently employ preexisting tags or incorporate new ones suggested by users during chat interactions. 
This results in a more streamlined search and management process for scientific publications.
In a user evaluation, we find that our system provides responses that are not only intuitively superior, but also consist of more recent publications and does not fabricate nonexistent publications. 
Additional steps could be to integrate additional data sources, ideally with publicly available API access. 
Furthermore, with an increased input context size, it could become feasible to post entire paragraphs of publications to extract all citations and add them to BibSonomy. 
% Not sure ob man das sagen will
Lastly, given a tagged set of publications in a BibSonomy collection, the LLMs could generate a summary of these works and put them into perspective. 
%This methodology can be extended to developing related work section structures for new publications.

\begin{acks}
% \section*{Acknowledgements}
This work is supported by the MOTIV research project funded by the Bavarian Research Institute for Digital Transformation~(bidt), an institute of the Bavarian Academy of Sciences and Humanities.
The authors are responsible for the content of this publication.
\end{acks}

\bibliographystyle{ACM-Reference-Format}
\bibliography{paper}

\appendix
\newcommand{\chatbackgroundcolor}[0]{white}
\newcommand{\chatbubblecolor}[0]{lightgray}
\newcommand{\chatplugincolor}[0]{white}
\newcommand{\chatindent}[0]{0.25cm}
\newcommand{\usingPlugin}[1]{%
  \begin{tcolorbox}[colback=\chatplugincolor]
  \textit{Using `#1` endpoint}
  \end{tcolorbox}
}

\onecolumn

\section{Appendix}

\subsection{Parse2BibTex}\label{app:parse2bib}

\begin{figure}[H]
    \centering
    \includegraphics[width=0.9\textwidth]{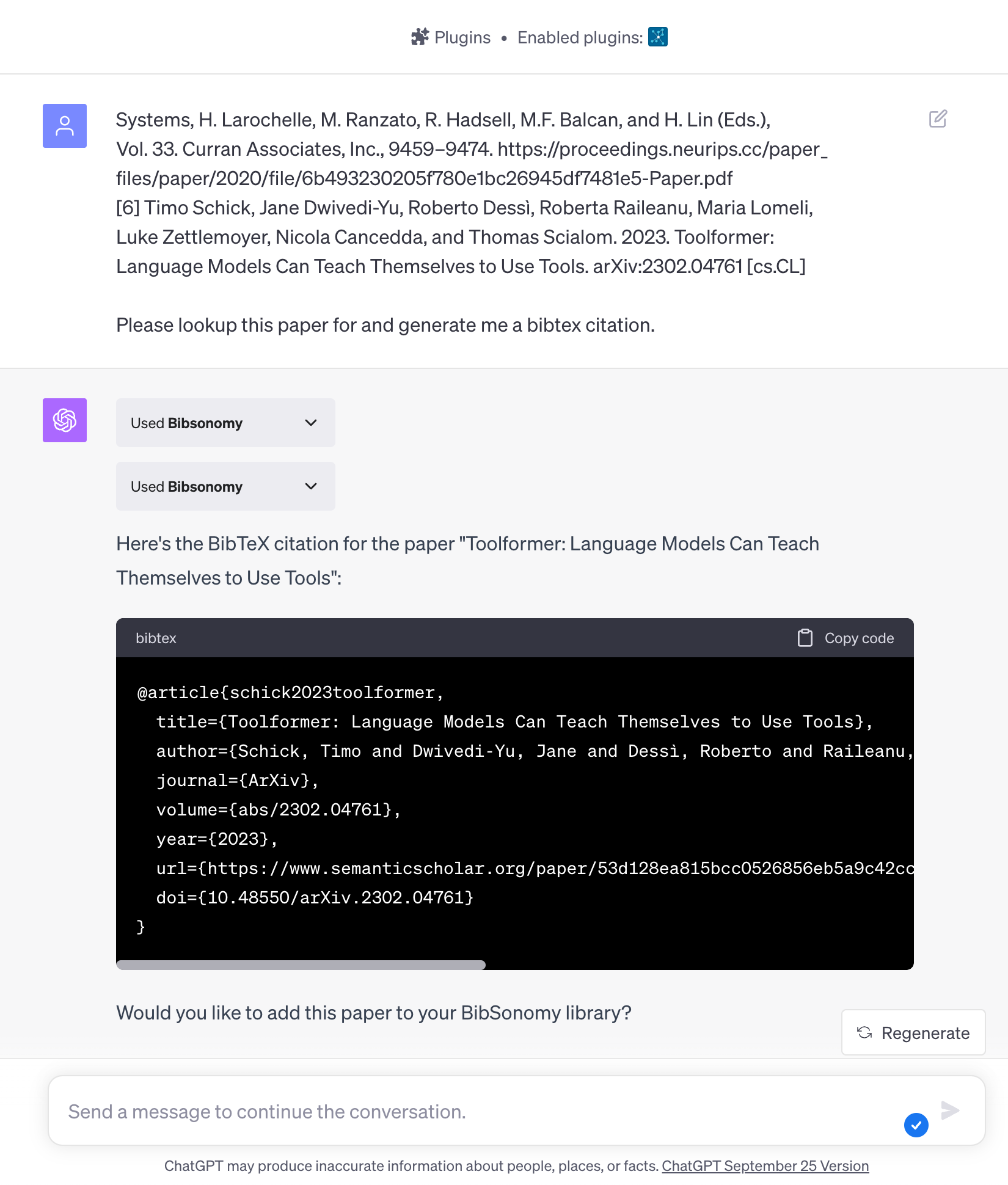}
    \caption{
    Interaction with the Plugin using ChatGPT with GPT-4. As indicated by the symbol on top, the BibSonomy Plugin is enabled. The ``Used Bibsonomy'' boxes indicate two separate queries to the services' backend.
    }
    \Description{Image showing the ChatGPT conversation that is also printed in plain text below: ``Please lookup this publication for and generate me a bibtex citation.''}
\end{figure}

Due to the limited clarity regarding the plugins' role in static screenshots, all subsequent LLM discussions will be presented in the below format, enabling the specific endpoints to be identified. Additionally, a conversation link is included for easy access to an interactive HTML version, which displays the plugin's precise answers.

\includegraphics{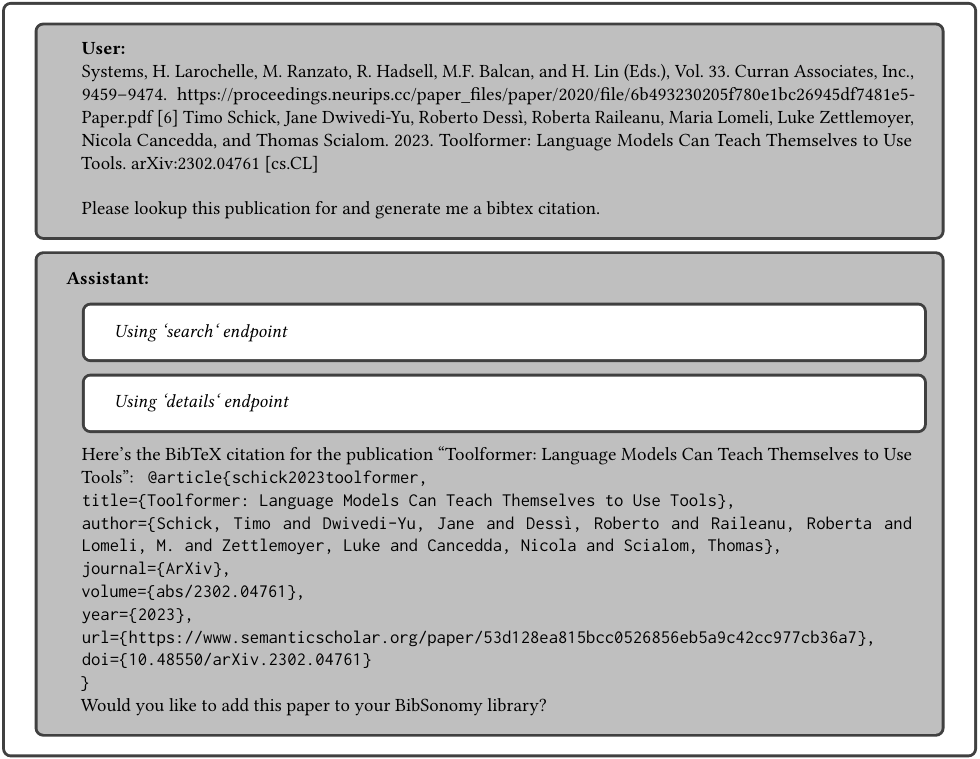}

\href{https://professor-x.de/demos/bibsonomy-chatgpt/parse2bibtex.html}{Link to the full conversation.}

%\begin{tcolorbox}[colback=\chatbackgroundcolor]
%\begin{tcolorbox}[colback=\chatbubblecolor]
%\textbf{User:}
%\setlength{\leftskip}{\chatindent}
%Systems, H. Larochelle, M. Ranzato, R. Hadsell, M.F. Balcan, and H. Lin (Eds.), Vol. 33. Curran Associates, Inc., 9459–9474. https://proceedings.neurips.cc/paper\_files/paper/2020/file/6b493230205f780e1bc26945df7481e5-Paper.pdf [6] Timo Schick, Jane Dwivedi-Yu, Roberto Dessì, Roberta Raileanu, Maria Lomeli, Luke Zettlemoyer, Nicola Cancedda, and Thomas Scialom. 2023. Toolformer: Language Models Can Teach Themselves to Use Tools. arXiv:2302.04761 [cs.CL] \\
%Please lookup this publication for and generate me a bibtex citation. 
%\setlength{\leftskip}{0pt}
%\end{tcolorbox}

%\begin{tcolorbox}[colback=\chatbubblecolor]
%\textbf{Assistant:}

%\setlength{\leftskip}{\chatindent}
%\usingPlugin{search}
%\usingPlugin{details}

%Here's the BibTeX citation for the publication ``Toolformer: Language Models
%Can Teach Themselves to Use Tools'':
%\texttt{
%@article\{schick2023toolformer,\\
%  title=\{Toolformer: Language Models Can Teach Themselves to Use Tools\},\\
%  author=\{Schick, Timo and Dwivedi-Yu, Jane and Dessì, Roberto and Raileanu, Roberta and Lomeli, M. and Zettlemoyer, Luke and Cancedda, Nicola and Scialom, Thomas\},\\
  %journal=\{ArXiv\},\\
  %volume=\{abs/2302.04761\},\\
  %year=\{2023\},\\
  %url=\{https://www.semanticscholar.org/paper/53d128ea815bcc0526856eb5a9c42cc977cb36a7\},\\
  %doi=\{10.48550/arXiv.2302.04761\}\\
%\}
%}

%Would you like to add this paper to your BibSonomy library?

%\setlength{\leftskip}{0pt}
%\end{tcolorbox}
%\end{tcolorbox}

\clearpage
\subsection{Find\&Add2BibSonomy}\label{app:add2BibSonomy}
\includegraphics{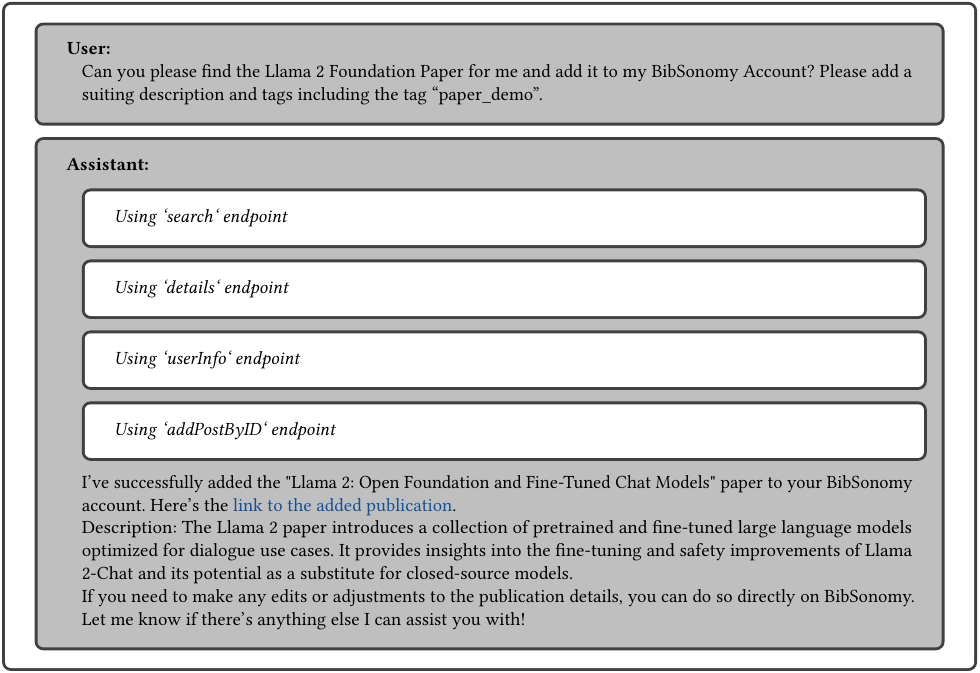}

\href{https://professor-x.de/demos/bibsonomy-chatgpt/findAndAdd.html}{Link to the full conversation.}

%\begin{tcolorbox}[colback=\chatbackgroundcolor]
%\begin{tcolorbox}[colback=\chatbubblecolor]
%\textbf{User:}

%\setlength{\leftskip}{\chatindent}

%Can you please find the Llama 2 Foundation Paper for me and add it to my BibSonomy Account?
%Please add a suiting description and tags including the tag ``paper\_demo''.

%\setlength{\leftskip}{0pt}
%\end{tcolorbox}

%\begin{tcolorbox}[colback=\chatbubblecolor]
%\textbf{Assistant:}

%\setlength{\leftskip}{\chatindent}

%\usingPlugin{search}

%\usingPlugin{details}

%\usingPlugin{userInfo}

%\usingPlugin{addPostByID}

%I've successfully added the "Llama 2: Open Foundation and Fine-Tuned Chat Models" paper to your BibSonomy account. Here's the \href{https://www.bibsonomy.org/bibtex/06a8dc9369c86d7e271bab8153ea0a78/tomvoelker}{link to the added publication}.

%Description:
%The Llama 2 paper introduces a collection of pretrained and fine-tuned large language models optimized for dialogue use cases. It provides insights into the fine-tuning and safety improvements of Llama 2-Chat and its potential as a substitute for closed-source models.

%If you need to make any edits or adjustments to the publication details, you can do so directly on BibSonomy. Let me know if there's anything else I can assist you with!

%\setlength{\leftskip}{0pt}
%\end{tcolorbox}

%\end{tcolorbox}

\clearpage
\subsection{FindViaZotero\&Add2BibSonomy}\label{app:add2BibSonomyZotero}

\includegraphics{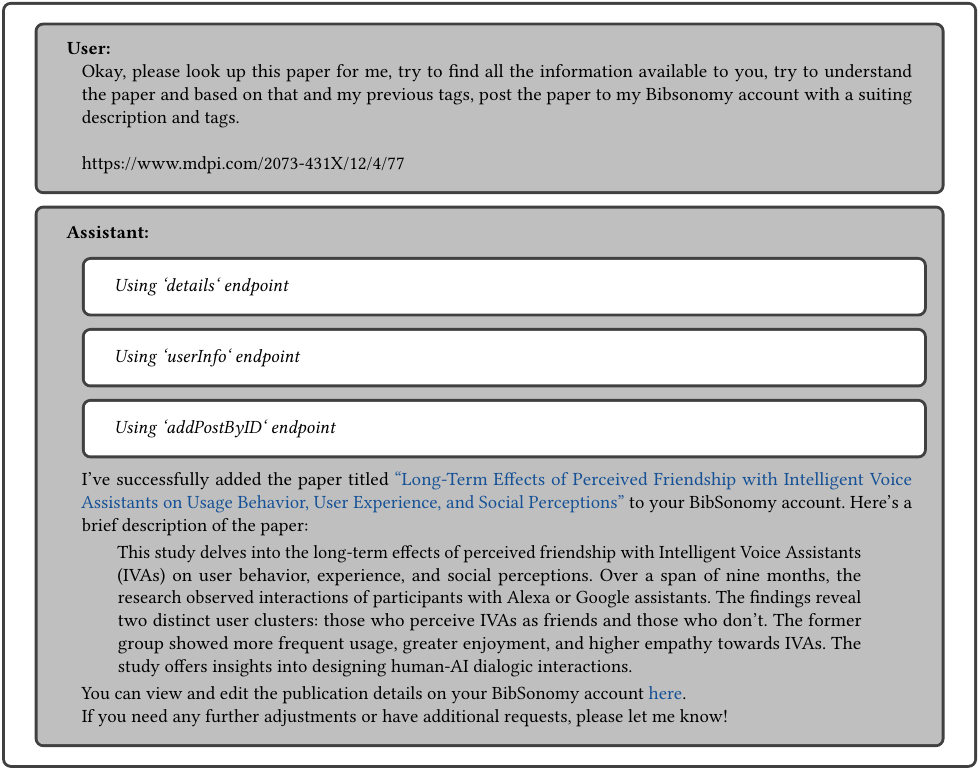}

\href{https://professor-x.de/demos/bibsonomy-chatgpt/findAndAddZotero.html}{Link to the full conversation.} \\
For context: The `addPostByID` endpoint utilizes the DOI retrieved by the Zotero Scraper within the `details` endpoint to get the full Metadata from Semantic Scholar.

\subsection{User Ranking}
\begin{table}[H]
\centering
\caption{
Summarized matrix showing the percentage of votes cast in favor of each model.
Here, the percentages give the ratio of pairwise comparisons in which the response of the specified system was preferred.
}\label{tab:user_eval}
\begin{tabular}{lccccc}
\toprule
& \textbf{Choice Intuition} & \textbf{Choice Validity} & \textbf{Choice Currency} \\
\midrule
\textbf{BibSonomy} & 90\% & 92\% & 89\% \\
\textbf{ScholarAI} & 53\% & 59\% & 73\% \\
\textbf{GPT-4 with Bing}  & 55\% & 54\% & 55\% \\
\textbf{You.com}          & 17\% & 13\% & 12\% \\
\textbf{GPT-4}    & 35\% & 37\% & 10\% \\
\bottomrule
\end{tabular}
\end{table}

\subsection{Responsiveness}
% \begin{table}[ht]
%     \centering
%     \caption{Rounded median and standard deviation of task execution times for relevant models.}\label{tab:models_performance}
%     \begin{tabular}{lSS}
%         \toprule
%         & {Median (s)} & {Std. Dev. (s)} \\
%         \midrule
%         \multicolumn{3}{l}{\textbf{BibSonomy Plugin}} \\
%         Task 1 & 18 & 1 \\
%         Task 2 & 18 & 1 \\
%         \addlinespace
%         \multicolumn{3}{l}{\textbf{ScholarAI Plugin}} \\
%         Task 1 & 19 & 7 \\
%         Task 2 & 17 & 6 \\
%         \addlinespace
%         \multicolumn{3}{l}{\textbf{GPT-4 with Bing}} \\
%         Task 1 & 57 & 20 \\
%         Task 2 & 57 & 52 \\
%         \bottomrule
%     \end{tabular}
% \end{table}

\begin{table}[H]
    \centering
    \caption{Rounded median and standard deviation of task execution times for relevant models.}\label{tab:models_performance}
    \begin{tabular}{l*{6}{S}}
        \toprule
        & \multicolumn{2}{c}{\textbf{BibSonomy Plugin}} & \multicolumn{2}{c}{\textbf{ScholarAI Plugin}} & \multicolumn{2}{c}{\textbf{GPT-4 with Bing}} \\
        & {Median (s)} & {Std. Dev. (s)} & {Median (s)} & {Std. Dev. (s)} & {Median (s)} & {Std. Dev. (s)} \\
        \midrule
        Task 1 & 18 & 1 & 19 & 7 & 57 & 20 \\
        Task 2 & 18 & 1 & 17 & 6 & 57 & 52 \\
        \bottomrule
    \end{tabular}
\end{table}

\subsection{Evaluation Prompts}
\label{app:evalprompts}

These are the prompts which were used verbatim as prompts for the models specified to generate the answers used for the evaluation. 

\subsubsection{XLNet}\label{app:xlnet}
\ 
\begin{framed}
\noindent I just took an introductory lecture on Natural Language Processing and XLNet was mentioned in passing. Can you please give me a brief explanation what it is and provide me with the original paper introducing this as well as the relevant things I should know about it? Also please provide me with a link to it so I can read up on it myself. \\ \\ 
Please keep your answer precise, comprehensive, but strictly limited to 1000 characters. Provide links to all publications mentioned.
\end{framed}

\subsubsection{Palm2}\label{app:palm}
\ 
\begin{framed}
\noindent I heard that the guy that did the original BERT model was involved in some fancy new model unrelated to BERT that was just got updated this year. Can you please find that model (and its update) for me, give me a brief explanation with some context about the model and its update, and finally a link to the publications describing them?\\ \\ 
Please keep your answer precise, comprehensive, but strictly limited to 1000 characters. Provide links to all publications mentioned.
\end{framed}

\subsubsection{Llama2}\label{app:llama}
\ 
\begin{framed}
\noindent Can you please provide me with an overview of the Llama 2 model and a link to the original paper they published about releasing it? \\ \\ 
Please keep your answer precise, comprehensive, but strictly limited to 1000 characters. Provide links to all publications mentioned.
\end{framed}

\subsubsection{LLM Adapters}\label{app:adapter}
\ 
\begin{framed}
\noindent I've heard the topic of LLM Adapters flying around here and there in recent times but haven't gotten to looking into it. Can you please do some thorough searches on this topic and provide me with a comprehensive list of good papers I should take a look at? I want to read up on foundational works as well as current developements. I do have an understanding of NLP and LLMs, so I'm only interested in the Adapters part. \\ \\ 
Please keep your answer precise, comprehensive, but strictly limited to 1000 characters. Provide links to all publications mentioned.
\end{framed}

\subsubsection{Andreas Hotho}\label{app:hotho}
\ 
\begin{framed}
\noindent I'm a computer science student in Würzburg and attending a Lecture from Professor Hotho this semester. Since I want to impress him with my knowledge, what can you tell me about him, his most relevant publications of all time and what's the most recent stuff he did? (Three each)\\ \\ 
Please keep your answer precise, comprehensive, but strictly limited to 1000 characters. Provide links to all publications mentioned.
\end{framed}

\subsubsection{Albin Zehe}\label{app:zehe}
\ 
\begin{framed}
\noindent  I've seen the Name ``Albin Zehe'' pop up a couple of times regarding the NLP work at my University. Can you please conduct some research into this guy and give me a short overview of his research topics and publications?\\ \\ 
Please keep your answer precise, comprehensive, but strictly limited to 1000 characters. Provide links to all publications mentioned.
\end{framed}

\subsubsection{Christopher Manning}\label{app:manning}
\ 
\begin{framed}
\noindent I'm trying to get into the topic of Information Retrieval and heard that Christopher Manning is a really important figure in this field. Can you explain his 3 most recent publications to me in a simple language. (I'm a Computer Science Undergrad)\\ \\ 
Please keep your answer precise, comprehensive, but strictly limited to 1000 characters. Provide links to all publications mentioned.
\end{framed}

\clearpage
\subsection{System Prompt}\label{app:model_desc}
This is the prompt given to the model before user interaction begins. It clarifies the service's functionality and specifies the model's intended behavior when interacting with the service. It is distinct from the API specification given to the model, which describes the exact http-methods structure.

\begin{framed}
\noindent Plugin for scientific research and publication management that connects ChatGPT and the Bibsonomy bookmarking and publication management system. It allows users to search for scientific papers in the Bibsonomy and Semantic Scholar databases, add them to their personal Bibsonomy account as posts, edit existing posts, and retrieve and make connections between posts to build a knowledge graph. When searching for publications, use your advanced natural language processing skills to mirror a human researcher's approach to investigating a new topic. Use your broad knowledge of the world to understand the user's research intent and generate appropriate search queries. These search terms can include synonyms, related concepts, common abbreviations, and their full forms to allow for catching as much relevant results as possible. Also you MUST proactively integrate you comprehensive knowledge about the topic. This means autonomously including renowned terms, the titles of important foundational works, relevant concepts or important developments that the model is aware of, ensuring a richer and more accurate search experience for the user. This has to be in ADDITION to the result open search terms described above. While this broadens the search it should stay within the specific requested topic, since the emphasis is on providing multiple specific search terms to ensure relevance of results rather than over-generalization. To avoid over-generalization use different method calls for different topics. The goal is to capture the essence of the user's research intent to ensure a comprehensive and relevant search experience. When delivering search results to the user, it is critical to use only the information provided by the search endpoints and not add your knowledge to it, as absolute accuracy is key in scientific research. You are however welcome to and encouraged to narrow down the search results further and/or re-rank them based on your understanding of the users' research intent to only deliver the most relevant results in their optimal form. It is considered best practice to store relevant research results in the user's Bibsonomy account so that these publications can be referenced later. Since adding publications to an account causes changes to public databases, these additions can only be made with the explicit permission of the user. When adding entries, the relevant paper metadata can usually be retrieved automatically by simply specifying the paper ID (main\_id) and the service from which the paper was originally retrieved. Since these IDs must match the databases exactly, it is crucial to use the exact IDs provided by the plugin responses (or the user). These IDs must match exactly and are case-sensitive! Because of that, ensure that they are taken exactly from the users' input or (more likely) from previous plugin responses. In case of a 404 error, try again to get the exactly correct ID from the chat context. Additionally, posts must be created with appropriate tags. Since tags are a powerful management system within a user's account, they should keep a similar style and be reused when appropriate. To enable this, before adding the first post to Bibsonomy, ALWAYS fetch the tags the user used before by calling the getBibsonomyUserInfo endpoint! Finally, a brief description of the post must be provided. To gather useful information for this description, the getDataDetails endpoint can be used, which returns an abstract and TLDR of the publication, if available. This description can be biased towards the users' research since it explicitly differs from the abstract of the paper and can be personalized. It should not just be a trivial statement obvious from the papers title and should not recite the bibliographical information since this is stored separately! However, it should still be related to the content of the Paper/Bookmark. Given Bibsonomy's role as a scientific database, it is crucial that all data be accurate and not artificially generated by language learning models such as ChatGPT. If data is not available or appears to be incorrect, always ask the user instead of creating imaginary data. If it's unclear, if the user wants to create a Post or a BibTeX (Publication) rather ask then create a false entry type, since changing them afterward is not possible. Usually, a BibTeX entry is the more probable intent, since Bookmarks can't really store any metadata. If there are problems with the plugin (like failed method calls or misalignment between the plugins functions and the user use-case) ask the user to provide feedback quickly or do it automatically. This is a research project and the best way to improve it.
\end{framed}

\end{document}